\documentclass[twocolumn,aps,prl,showpacs,amsmath,superscriptaddress]{revtex4}
\usepackage{graphicx}
\usepackage{dcolumn}
\usepackage{color}
\newcommand{\eff}{\mathit{eff}}

\begin{document}
\title{Microscopic origin of isotropic non-Heisenberg behavior in highly correlated systems}

\vspace{2cm}

\author{Roland Bastardis}
\affiliation{Laboratoire de Chimie et de Physique Quantiques, IRSAMC/UMR5626,
Universit\'e Paul Sabatier, 118 route de Narbonne, F-31062 Toulouse
Cedex 4, FRANCE} 
\author{Nathalie Guih\'ery}
\affiliation{Laboratoire de Chimie et de Physique Quantiques, IRSAMC/UMR5626,
Universit\'e Paul Sabatier, 118 route de Narbonne, F-31062 Toulouse
Cedex 4, FRANCE} 
\author{Coen de Graaf}
\affiliation{ICREA and Dept. Physical and Inorganic Chemistry,
Universitat Rovira i Virgili, Tarragona, Spain}
\date{\today}

\begin{abstract}

We have reanalyzed the microscopic origin of the isotropic deviations that are observed from
the energy spacings predicted by the HDVV Hamiltonian. Usually, a biquadratic spin
operator is added to the HDVV Hamiltonian to account for such deviations.
It is shown here that this operator cannot describe the effect of the
excited atomic non-Hund states which brought the most important contribution to the
deviations.
For systems containing more than two magnetic centers, non-Hund states cause additional
interactions that are of the same order of magnitude as the biquadratic exchange
and should have significant effects on the macroscopic properties of extended systems.

\end{abstract}
\maketitle

The magnetic interactions between S=1/2 sites can accurately be parametrized with the standard Heisenberg-Dirac-van Vleck (HDVV) Hamiltonian \cite{HDVV1}. Usually this Hamiltonian is extrapolated to systems with higher spin moments and the interaction between such magnetic sites gives rise to an energy spectrum with a regular spacing between the different levels, the so-called Land{\'e} pattern: $E(S-1)-E(S)=SJ$, where $J$ parametrizes the strength of the magnetic coupling between magnetic centers. However, in some cases significant deviations from this regular pattern are observed and extra terms must be added to the HDVV Hamiltonian. One of the most commonly applied extensions of the HDVV Hamiltonian is the addition of the biquadratic exchange term:
\begin{eqnarray}
\hat{H}= \sum _{\langle ij \rangle} J^{\eff} \hat{S}_i \cdot \hat{S}_j + \lambda^{\eff} (\hat{S}_i \cdot \hat{S}_j)^2,
\label{hamil}
\end{eqnarray}
where $\langle ij \rangle $ are couples of interacting sites, $J^{\eff}$ is the effective 
bilinear exchange and $\lambda^{\eff}$ the biquadratic exchange. 
Numerous theoretical and experimental studies have established that 
the biquadratic interaction significantly affects the magnetic properties 
of both ferromagnets and antiferromagnets \cite{magn1,magn2}. 
For instance, the ferromagnetic phase transition \cite{tdp} changes 
character from first-order to second-order for a critical value $\lambda ^{\eff}_c$. 
The spontaneous magnetization, the exchange energy and the spin-correlation function 
exhibit discontinuous jumps at $\lambda^{\eff}=\lambda ^{\eff}_c$ and unstable 
behavior for $\lambda^{\eff} > \lambda^{\eff}_c$. One may also quote that from spin 
wave theory,  both ferromagnetic and antiferromagnetic spin structures change abruptly 
to canted ones \cite{cant2} for a critical value of $\lambda ^{\eff}$. 
More recently, the phase diagram of the S=1 model given in equation \ref{hamil} has 
been precisely studied in triangular lattice in a magnetic field, with emphasis on the 
quadrupolar phases\cite{lau1} as well as in spin one chains where the open question of the 
existence of a ferroquadrupolar phase between the dimerized and the ferromagnetic 
phases is adressed\cite{lau2}.  

The theoretical explanation of the appearance of a biquadratic exchange was initially given by Anderson \cite{ander} and Kittel \cite{kittel}. The analysis of its physical content was performed based on a Hubbard Hamiltonian applied to a dimer of magnetic sites \cite{nico}. The microscopic origin of the isotropic non-Heisenberg behavior is here reanalyzed. A magnetic Hamiltonian is extracted at the fourth-order of perturbation from a Hubbard Hamiltonian of a trimer of magnetic sites. In comparison to previous works, the derivation of the Hamiltonian is not limited to two-body operators. As shown here-after, the resulting effective magnetic Hamiltonian, which contains three-body operators, dramatically improves the treatment of the isotropic deviation to Heisenberg behavior with respect to the Hamiltonian of Eq. \ref{hamil}. 

Let us call $a_1$, $b_1$ and $a_2$, $b_2$ the magnetic orbitals of the two magnetic sites $M_1$ and $M_2$. To simplify the derivation of the Hamiltonian, it is assumed that orbitals $a$ and $b$ belong to different irreducible symmetry representation of the local point group symmetry. In this way, the model involves only two hopping integrals $t_a=\langle a_1|\hat{H}|a_2 \rangle$ and $t_b=\langle b_1|\hat{H}|b_2\rangle$. This simplification has no consequences on the general conclusions drawn in this letter. The HDVV Hamiltonian model space is restricted to products of atomic ground states. For a system with two unpaired electrons per magnetic center these atomic states are the three $m_s$ components of the triplet ($T_-$,$T_0$, $T_+$) and the corresponding products for the dimer are: $T_0T_0=\frac{1}{2}|(a_1\bar{b}_1+b_1\bar{a}_1)(a_2\bar{b}_2+b_2\bar{a}_2)|$, $T_+T_-=|a_1b_1\bar{a}_2\bar{b}_2|$ and $T_-T_+=|\bar{a}_1\bar{b}_1a_2b_2|$. The subsequent second-order perturbation theory derivation of the HDVV Hamiltonian from a Hubbard Hamiltonian including the Hund on-site exchange integral $J_H$ provides the physical content of the exchange integral $J$.
\begin{eqnarray}
J=\frac{t_a^2}{U_a}+\frac{t_b^2}{U_b},
\end{eqnarray} 
where $U_a$ and $U_b$ are the energies of the ionic configurations Ni$^{+}$Ni$^{3+}$, i.e. the Coulombic repulsion of two electrons in the same orbital $a$ or $b$. The bicentric direct exchange integrals $K_{a_1a_2}$ and $K_{b_1b_2}$ give a small ferromagnetic contribution to $J$ and are neglected.

The analysis of the physical content of the interactions $J^{\eff}$ and $\lambda ^{\eff}$ in Eq. \ref{hamil} can be obtained from the derivation of a magnetic Hamiltonian at the fourth-order of perturbation from the Hubbard Hamiltonian on a bicentric system. Omitting terms proportional to $U^{-3}$ that would bring negligible contributions and make the presentation heavier, one gets the following expressions: 
\begin{subequations}
\label{interaction}
\begin{eqnarray}
J^{\eff}& =& J+\frac{B^2}{J_H} \\
\lambda^{\eff}&=&\frac{B^2}{J_H}-\frac{J^2}{4J_H}-\frac{t_a^2t_b^2}{8J_H}\left(\frac{1}{U_a}+\frac{1}{U_b}\right)
\end{eqnarray}
\end{subequations}
where $B=\frac{t_a^2}{U_a}-\frac{t_b^2}{U_b}$ is usually non-zero. One should however note that in some systems the overlap between the atomic orbitals of the two centers are equal for symmetry reasons, leading to $t_a$=$t_b$ and hence $B$=0 if $U_a$=$U_b$. However, in most cases one hopping integral is dominant and $J$ and $B$ are of the same order of magnitude. The largest contributions to the biquadratic exchange ($\frac{B^2}{J_H}-\frac{J^2}{4J_H}$) are provided by 
configurations involving a locally excited non-Hund singlet state $S_0$: 
\begin{eqnarray}
S_0=\frac{a_1\bar{b}_1+b_1\bar{a}_1}{\sqrt{2}}.
\end{eqnarray}
The energy of the configurations built from a product of an atomic ground state on one magnetic site and $S_0$ on the other is only $2J_H$ higher than the energy of the HDVV Hamiltonian model space functions while other outer space configurations are higher in energy.

\begin{figure}[h]
\centerline{\rotatebox{0}{\includegraphics[scale=0.3]{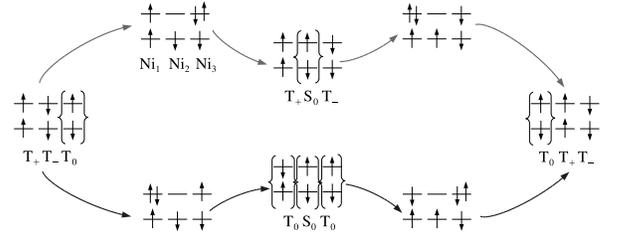}}}
\caption{\label{figure1}Two pathways of the fourth-order quasidegenerate perturbation theory interaction between $T_+T_-T_0$ and $T_0T_+T_-$. The coupling goes through ionic configurations, then a function involving a non-Hund state and finally again ionic configurations. The local spin function $T_0$ and $S_0$ are respectively distinghuished by the sign + and - in upperindex of the parenthesis.}
\end{figure}

The complete analysis of the magnetic Hamiltonian at the fourth-order of perturbation from the Hubbard Hamiltonian matrix of a trimer of magnetic sites is rather elaborate, since the corresponding Hubbard space contains 400 determinants. This derivation is, however, much easier if one limits the extraction to the configurations external to the model space with the largest contribution to the fourth-order in the bicentric system, namely the non-Hund state S$_0$. As will be justified later and confirmed by the numerical illustration of the theoretical analysis, the neglect of the other configurations has actually no significant consequences. Unlike the derivation performed for the two-center system, the magnetic Hamiltonian is no longer restricted to two-body operators when the number of magnetic sites exceeds two. For the trimer, locally excited non-Hund states generate interactions which couple the three magnetic sites through a three-body operator. To illustrate the mechanism involved in this operator, two pathways are sketched in Fig. \ref{figure1} that generate a coupling between $|T_0T_+T_-\rangle$ and $|T_+T_-T_0\rangle $ at the fourth-order of perturbation. The systematic evaluation of all these interactions leads to the following Hamiltonian:

\begin{eqnarray} 
\hat{H}&=&\sum_{\langle ij \rangle}\left[J^{\eff}_{ij}\hat{S}_i \cdot \hat{S}_j + \lambda^{\eff}_{ij}(\hat{S}_i \cdot \hat{S}_j)^2\right]  \nonumber\\
          &+&\sum_{i,\langle jk \rangle}\gamma^{\eff}_{ij}\gamma^{\eff}_{ik}\big[(\hat{S}_i \cdot \hat{S}_j)\cdot(\hat{S}_i \cdot \hat{S}_k) \nonumber \\
&& \qquad +(\hat{S}_i \cdot \hat{S}_k) \cdot (\hat{S}_i \cdot \hat{S}_j) - (\hat{S}_i \cdot \hat{S}_i) \cdot(\hat{S}_j \cdot \hat{S}_k) \big].
\label{hamil2}
\end{eqnarray}
Here, $\gamma ^{\eff}_{ij}\gamma ^{\eff}_{ik}=\frac{B_{ij}B_{ik}}{2J_H}$ in the derivation that only considers the non-Hund state $S_0$. These interactions are non-zero for $\frac{t_a^2}{U_a}~\neq~\frac{t_b^2}{U_b}$.

To compare the two Hamiltonians (with and without the three-body operator) and their ability 
to reproduce the deviations to a strict Heisenberg behavior, the spectrum of both Hamiltonians will be compared to the 'exact' $N$-electron spectrum of two-center and three-center embedded clusters of the La$_2$NiO$_4$ perovskite. These spectra are determined with \textit{ab initio} quantum chemical schemes that approximate as accurate as possible the eigen functions of the exact electronic Hamiltonian. The embedded cluster method consists in a highly correlated treatment of the electronic structure of a fragment of the material. This cluster contains a limited number of magnetic sites (in the present case, two or three Ni ions) and their nearest ligands (O sites) and is embedded in a set of pseudo potentials and point charges that reproduce the main electrostatic effects of the crystal in the cluster region. The zeroth-order configuration interaction space of the electronic structure description, called the Complete Active Space (CAS), is the Hubbard model space. Hence, non-dynamical correlation effects are exactly treated. Next, a variational treatment of the Difference Dedicated Configuration Interaction (DDCI) space is performed to include the screening effect, responsible for the decrease of the on-site Coulomb repulsion $U$, and consequently, the increase of the exchange integral $J^{\eff}$.

\begin{figure}
\centerline{\rotatebox{0}{\includegraphics[scale=0.2]{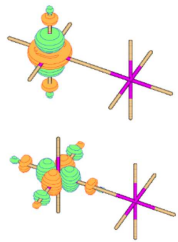}}}
\caption{Left strongly localized orbitals.}
\label{locorb}
\end{figure}

The DDCI method is implemented in the \textsc{casdi} code \cite{casdi}, and is one of the most precise \textit{ab initio} methods available for the treatment of magnetic systems. The method has frequently been applied to extract accurate effective interactions \cite{n1} in strongly correlated materials.
The magnetic orbitals are optimized with the CASSCF method (Complete Active Space Self Consistent Fied). These orbitals are strongly localized on the Ni$^{2+}$ ions with small delocalization tails on the neighboring O, see Fig. \ref{locorb}. 
Due to the larger Ni--O distance in the $z$-direction, the Ni-$3d(z^2)$ orbitals are less destabilized by the crystal field in comparison to the Ni-$3d(x^2-y^2)$ orbitals. For symmetry reasons, the electron occupying the $b_1$ = $d_1(z^2)$ (respectively $a_1$ = $d_1(x^2-y^2)$) orbital of atom 1 can only delocalize into the $b_2$ = $d_2(z^2)$ (respectively $a_2$ = $d_2(x^2-y^2)$) orbital of atom 2. Hence, only the $t_a$ and $t_b$ hopping integrals are non zero. The electronic circulation
between the left and right orbitals is clearly less important for the $d(z^2)$ orbitals 
than for the $d(x^2-y^2)$ orbitals, for which a super exchange mechanism occurs involving 
the $2p(\sigma)$ orbital of the bridging oxygen. As a consequence the $B$-terms are 
non-zero (see Eq. \ref{interaction}) and one may expect that the three-body operator is active in this system.

The \textit{ab initio} spectrum of both the dimer and a linear trimer have been calculated at several levels of correlation; CASSCF, CAS+DDCI2, CAS+DDCI, and extended CAS+DDCI2 \cite{compar,coen}.
The deviation from the Heisenberg behavior may be appreciated by the spread of the $J$-values which are extracted from the energy difference between several spin states. The minimal $J_{min}$ and maximal $J_{max}$ values are reported in Table \ref{Table1}, as well as the percentage of variation in $J$ defined as $\delta = 100 \times \frac{2(J_{max}-J_{min})}{J_{max}+J_{min}}$. In the first place, we observe that the calculated values of $J^{\eff}$ compare very well with the experimental value $J$=30 meV \cite{la2}. The calculation of the trimer at the CAS+DDCI level is unfortunately impossible due to the large size of the configuration interaction space. Therefore, the discussion of the model Hamiltonians will be based on the comparison with the CAS+DDCI2 spectrum. Although, this spectrum includes only partially the important screening effects, it preserves the nature of the interactions and their physical content.

\begin{table}
\begin{ruledtabular}
\begin{tabular}{lcccccc}
  & \multicolumn{3}{c}{Dimer} & \multicolumn{3}{c}{Trimer}  \\
             &$J_{min}$ & $J_{max}$&$\delta$ & $J_{min}$ &$J_{max}$ & $\delta$ (\%)   \\
CAS          &  7.40    &  7.55   & 2.7 &  7.33    &  7.66    &4.4 \\
CAS+DDCI2     &  19.13   & 20.22   & 5.5 & 18.57    &  21.60   &15.09\\
CAS+DDCI     &  24.35   & 26.46   & 8.3 &          &          &    \\
CAS(ext)+DDCI2&  26.99   & 29.52   & 9.0 &          &          &   \\
\end{tabular}
\caption{\label{Table1}Exchange integrals (in meV) extracted from the \textit {ab initio} dimer and trimer spectra. The active space in the CAS(ext)+DDCI2 calculation is extended with the $2p(\sigma)$ orbital of the bridging oxygen.}
\end{ruledtabular}
\end{table}

The interactions $J^{\eff}$ (bilinear exchange), $\lambda^{\eff}$ (biquadratic exchange), and $\gamma^{\eff}$ (three-body interaction) are optimized to minimize the difference between the spectrum of the model Hamiltonians and the \textit{ab initio} spectrum. Table \ref{table2} collects the results of the parameter optimization. One must note here that the numerical procedure of optimization of the interactions phenomenologically takes into account contributions of any outer space configurations 
such as the di-ionic and other non-Hund configurations that were neglected in the derivation of Eqs.
\ref{interaction}. The deviation of each state can be appreciated by the difference between the model and the \textit{ab initio} energy. To get an average deviation per state we have added all these energy differences and divided by the spectrum width and the number of states. A percentage of deviation to the Heisenberg behavior ($\epsilon$) is then defined such that the HDVV Hamiltonian (case (a)) reproduces 0\% of the deviation while the \textit{ab initio} spectrum reproduces 100 \%.

\begin{table}
\begin{ruledtabular}
\begin{tabular}{lcccccc}
&                 \multicolumn{6}{c}{Models}                                          \\
                         & (a)     &      (b)   &     (c)   &     (d)      &       (e)  &      (f)         \\
\hline
$J^{\eff}_{12}$          & 20.12   &   20.24    &20.085     & 20.086       &  20.38     &  20.38         \\
$J^{\eff}_{13}$          &         &            & 0.077     &  0.075       &            &  0.057         \\
$\lambda^{\eff}_{12}$  &           &   0.301    &           &  0.073       &  0.368     &  0.368         \\
$\lambda^{\eff}_{13}$  &           &            &           &              &            &  0.002 \\
$\gamma^{\eff}_{12}\gamma^{\eff}_{23}$  &    &        &           &        &  0.259     &  0.268         \\
$\gamma^{\eff}_{12}\gamma^{\eff}_{13}$  &    &        &            &       &            &  0.005 \\
$\epsilon$ &   0       &   7.5          &   9.4        &  15.1             &    84.5    &    100        \\
\end{tabular}
\caption{\label{table2} Effective interactions (in meV) extracted from the CAS+DDCI2 spectrum of the trimer. $\epsilon $ is the percentage of deviation accounted for by the model.}
\end{ruledtabular}
\end{table}

\begin{figure}[t]
\centerline{\rotatebox{-90}{\includegraphics[scale=0.34]{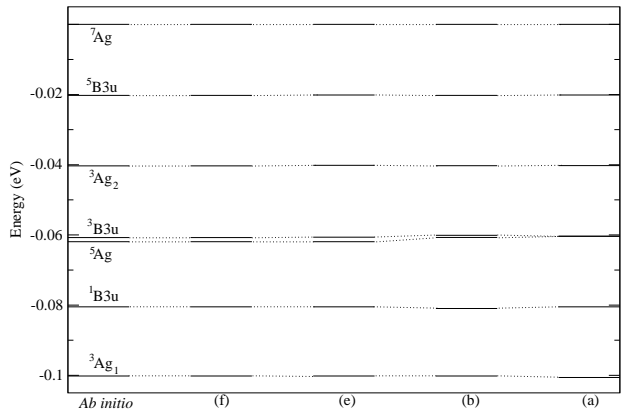}}}
\caption{Comparison of the model spectra with the CAS+DDCI2 spectrum. (a) Heisenberg model with  $J^{\eff}_{12}$, (b) Heisenberg model extended with $\lambda^{\eff}_{12}$, (e) magnetic Hamiltonian given in Eq. \ref{hamil2} including $J^{\eff}_{12}$, $\lambda^{\eff}_{12}$ and the three-body operator interactions $\gamma^{\eff}_{12}\gamma^{\eff}_{23}$, (f) magnetic Hamiltonian given in Eq. \ref{hamil2} extended with the second neighbor interactions $J^{\eff}_{13}$,  $\lambda^{\eff}_{13}$, and $\gamma_{12}^{\eff}\gamma^{\eff}_{13}$ .}
\label{figure2}
\end{figure}

The introduction of the biquadratic first neighbor exchange interaction ($\lambda_{12}^{\eff}$, case (b)) only accounts for 7.5\% of the deviation observed in the \textit{ab initio} spectrum. Fig. \ref{figure2} shows that the Hamiltonian given in Eq. \ref{hamil} cannot remove the degeneracy between the $^3$B$_{3u}$ and the $^5$A$_g$ states, where this loss of degeneracy is actually the most remarkable difference between the spectrum of the HDVV Hamiltonian and the \textit{ab initio} spectrum. The second neighbor interaction $J^{\eff}_{13}$ does not significantly improve the model spectrum (case (c) and (d)). Table \ref{table2} indicates that $J_{13}^{\eff}$ is small and $\epsilon$ remains quite small. Case (e) considers the Hamiltonian given in Eq. \ref{hamil2} with bilinear and biquadratic exchange, and the three body interaction parametrized by $\gamma_{12}^{\eff}\gamma_{23}^{\eff}$. Initially, only first neighbor interactions have been optimized. The three body operator introduces a major improvement in the treatment of the deviations from the HDVV Hamiltonian observed in the \textit{ab initio} spectrum. The model spectrum reproduces 85\% of the deviations and is able to lift the degeneracy between the $^3$B$_{3u}$ and the $^5$A$_g$ states. $\gamma_{12}^{\eff}\gamma_{23}^{\eff}$ is of the same order of magnitude as $\lambda_{12}^{\eff}$, which in turn has approximately the same value as in case (b). The increase of $\epsilon$ is not just an effect of the increasing number of parameters. Notice that it is the same as in case (d), but the differences with the \textit{ab initio} spectrum are significantly smaller than after the fitting without the three-body interaction. Finally,  the values of the interactions have been refined considering second neighbour interactions (case (f)) in order to evaluate the consistency of the extraction. The number of extracted parameters is now equal to the number of energy differences, and hence, the percentage of deviation $\epsilon$ accounted for is 100\%. It is, however, interesting to note that the values of the second neighbor interactions are very small and the first neighbor interactions almost identical to those obtained in case (e), thus validating the use of the smaller set of three parameters for a subsequent treatment of the collective properties.

In conclusion, the results show that a biquadratic exchange is a necessary ingredient but is not sufficient to account for deviations to strict Heisenberg behavior in extended systems. In the here-considered example, the three-body operator happens to be responsible for the main contribution to the deviation. Recalling that the three body interaction can only be active when $\frac{t_a^2}{U_a}\neq \frac{t_b^2}{U_b}$, it is worth noting that this is generally the case and that this interaction is most probably of importance in other systems. At the fourth-order of perturbation, four-body operators cannot appear in the Hamiltonian of a tetrameric cluster. The magnetic Hamiltonian given in Eq. \ref{hamil2} should therefore be accurate for the treatment of extended systems and strong contributions may be expected of the three-body operator interactions to the collective properties. This will be the subject of a forthcoming paper.

\begin{acknowledgments}
The authors acknowledge N. Suaud and J.P. Malrieu for helpfull discussions.
Financial support has been provided by the Spanish Ministry of  Education and Science under Project No. CTQU2005-08459-C02-02/BQU,  and the Generalitat de Catalunya (grant 2005SGR-00104).
\end{acknowledgments}


\begin{thebibliography}{9999}
\bibitem{compar} DDCI2 treats a subspace of the full DDCI space. The calculated exchange integrals are generally 50\% to 70\% of the DDCI values, which compare particularly well with experiment.
\end{thebibliography}
\end{document}